\documentclass[a4paper,11pt]{article}
\usepackage{pos}
\usepackage{float}
\usepackage{graphicx}
\usepackage{url}
\usepackage{booktabs}
\usepackage{siunitx}
\usepackage{comment}
\usepackage{upgreek}
\usepackage{lineno}

\newcommand{\novbb}{$0\nu\beta\beta$}
\newcommand{\twovbb}{$2\nu\beta\beta$}
\newcommand{\qbb}{Q$_{\beta\beta}$}
\newcommand{\mbb}{m$_{\beta\beta}$}
\newcommand{\tild}{~}

\title{First results from LEGEND-200: searching for $0\nu\beta\beta$ decay in $^{76}$Ge}

\author[a,b,c]{Giovanna Saleh}
\onbehalf{on behalf of the LEGEND collaboration}

\affiliation[a]{Dipartimento di Fisica e Astronomia, Università degli Studi di Padova,\\
Via Marzolo 8, 35131, Padua, Italy}
\affiliation[b]{Istituto Nazionale di Fisica Nucleare, Sezione di Padova, \\
Via Marzolo 8, 35131, Padua, Italy}
\affiliation[c]{Physik-Institut,  Universität Zürich, \\
Winterthurerstrasse 190, 8057, Zürich, Switzerland}

\emailAdd{giovanna.saleh@phd.unipd.it}
\emailAdd{giovanna.saleh@physik.uzh.ch}

\abstract{
	The LEGEND Experiment searches for the neutrinoless double beta (0$\nu\beta\beta$) decay of $^{76}$Ge employing active high purity germanium detectors enriched in $^{76}$Ge beyond 86\%. LEGEND's experimental program is articulated in two phases: LEGEND-200, currently ongoing, and LEGEND-1000, the next generation development.
	LEGEND-200 started operating in 2023 at Laboratori Nazionali del Gran Sasso (LNGS) and ran in a stable physics data taking regime for about one year with 142.5 kg of detectors installed. 
	With a target background index of $2 \cdot 10^{-4} $ counts/(keV kg yr) at Q$_{\beta\beta} \sim$ 2039 keV and a final exposure of 1000 kg yr, LEGEND-200 aims to reach a 3$\sigma$ discovery sensitivity for a 0$\nu\beta\beta$ half-life of  $10^{27}$ yr.\\
	In this contribution, the LEGEND-200 experiment will be presented, with a focus on its current status and on the results obtained with the first year of data.
	In particular, the employed analysis routines will be introduced, the signal identification and background suppression performance will be discussed, and the background appearing in the region of interest around $Q_{\beta\beta}$ will be analyzed.
	The performed analysis of LEGEND-200 data finds no evidence for a \novbb{} signal: a lower limit to its half-life is set instead, $T_{1/2}^{0\nu} > 0.5 \cdot 10^{26}$ yr, at 90\% CL. A joint GERDA + MAJORANA Demonstrator + LEGEND-200 analysis provides a limit of  $T_{1/2}^{0\nu} > 1.9 \cdot 10^{26}$ yr, at 90\% CL\tild\cite{prl25}. \\
	
	This work is supported by the U.S. DOE and the NSF, the LANL, ORNL and LBNL LDRD programs; the European ERC and Horizon programs; the German DFG, BMBF, and MPG; the Italian INFN; the Polish NCN and MNiSW; the Czech MEYS; the Slovak RDA; the Swiss SNF; the UK STFC; the Canadian NSERC and CFI; the LNGS and SURF facilities.

}

\FullConference{The European Physical Society Conference on High Energy Physics (EPS-HEP2025)\\
7-11 July 2025\\
Marseille, France\\}

\begin{document}
\maketitle
\raggedbottom

\section{Introduction}

Neutrinoless double beta (\novbb) decay offers the possibility of directly testing lepton number conservation and the Dirac/Majorana nature of neutrinos. In fact this process, in which two $\beta$ decays occur simultaneously in a nucleus, ($A,Z) \rightarrow (A, Z+2) + 2e^-$, is not allowed within the current formulation of the Standard Model (SM), can take place only if neutrinos are Majorana fermions and implies a lepton number violation of two units ($\Delta L =2$). The observation of the \novbb{} decay would also provide some insights into the matter-antimatter asymmetry puzzle, as it would support some leptogenesis mechanism, which also requires neutrinos to be Majorana fermions.

In the hypothesis of the process being mediated by light Majorana neutrinos exchange, the rate of the \novbb{} decay is proportional to the square of the effective Majorana mass of the neutrinos, $m_{\beta\beta}=\sum_i \big|U_{ei}^2m_i\big|$, according to the relation $[T_{1/2}^{0\nu}]^{-1} = G^{0\nu}(Q_{\beta\beta}, Z) |M^{0\nu}|^{2} \frac{|m_{\beta\beta}|^{2}}{m_e^2}$, in which $G^{0\nu}$ is the phase space integral, $M^{0\nu}$ the nuclear matrix element and $m_e$ the mass of the electron \tild\cite{disc_matter_antimatter}. An experimental measurement of (or a limit on) the half-life translates then to an estimate of (or a limit on) the effective Majorana mass, which would give some information about the sum of the neutrino masses. The allowed parameters space for \mbb{} strongly depends on the hypothesis on the neutrino mass ordering: for the Inverted Ordering (IO) \mbb{} is expected to be > 10 meV\tild\cite{benato}, while for the Normal Ordering (NO) \mbb{} can be arbitrarily small due to potential cancellations between terms.
Fully probing the IO parameters space is a major goal of the next-generation \novbb{} experiments, including LEGEND-1000. 

In the SM counterpart of \novbb{}, called two neutrinos double beta (\twovbb) decay, the two emitted electrons appear in the final state together with two neutrinos: ($A,Z) \rightarrow (A, Z+2) + 2e^- + 2\bar{\nu}_{e}$. In \twovbb{} the energy of the decay is therefore shared between the electrons and the neutrinos, resulting in a continuous electron sum energy spectrum with energies ranging from 0 to the total available energy, corresponding to the Q-value of the decay (\qbb). The neutrinoless mode, instead, would result in the electrons carrying the total energy of the decay: the experimental signature of \novbb{} is therefore a narrow peak at \qbb{} in the electron sum energy spectrum.

Different $\beta\beta$-decaying isotopes can be used to search for \novbb{}: common selection criteria include a high isotopic abundance, a high \qbb{} to minimize the natural background in the region of interest and the compatibility with scalable experimental techniques.

\section{The LEGEND project}

The Large Enriched Germanium Experiment for Neutrinoless $\beta\beta$ Decay (LEGEND) searches for \novbb{} in $^{76}$Ge (\qbb{} = $2039.061 \pm 0.007$ keV\tild\cite{qbb_value}).
It uses active High Purity Germanium (HPGe) detectors enriched in $^{76}$Ge, serving both as the source of the $\beta\beta$ decaying isotope and as detectors to measure the energy of the emitted electrons.

The major advantages of using germanium are its excellent energy resolution, reaching a FWHM(\qbb) $\sim$ 2 keV (0.1\%) and its low intrinsic background level.
In fact, for $\beta\beta$ experiments it is important to operate in a background-free regime, in which less than one background count is expected in the region of interest over the full duration of the data taking: in this regime the half-life discovery sensitivity (DS) scales linearly with the exposure $\varepsilon = M \cdot t$, rather than with its square root, which is the case in the background-dominated regime. This is reported in Eq.~\ref{hlsens}, in which $M$ is the active mass, $t$ the data taking time, $\sigma$ the energy resolution at \qbb, and $B$ the background level.
\begin{equation}
	T_{1/2}^{0\nu} \,  \text{ DS} \propto 
	\begin{cases}
		M \cdot t & \text{Background-free} \\
		\sqrt{\frac{M \cdot t}{B \cdot \sigma}} & \text{With background}
	\end{cases}
	\label{hlsens}
\end{equation}

The LEGEND Project foresees a staged approach to ultimately achieve a quasi-background-free regime (Fig.\tild\ref{discovery_sensitivity}) and a half-life sensitivity beyond $10^{28}$ years (Table\tild\ref{tab:legend_project}).

\begin{minipage}{0.44\textwidth}
	\begin{figure}[H]
		\hspace*{-0.9cm}\parbox{\dimexpr\textwidth+0.9cm\relax}{%
			\includegraphics[width=1\textwidth]{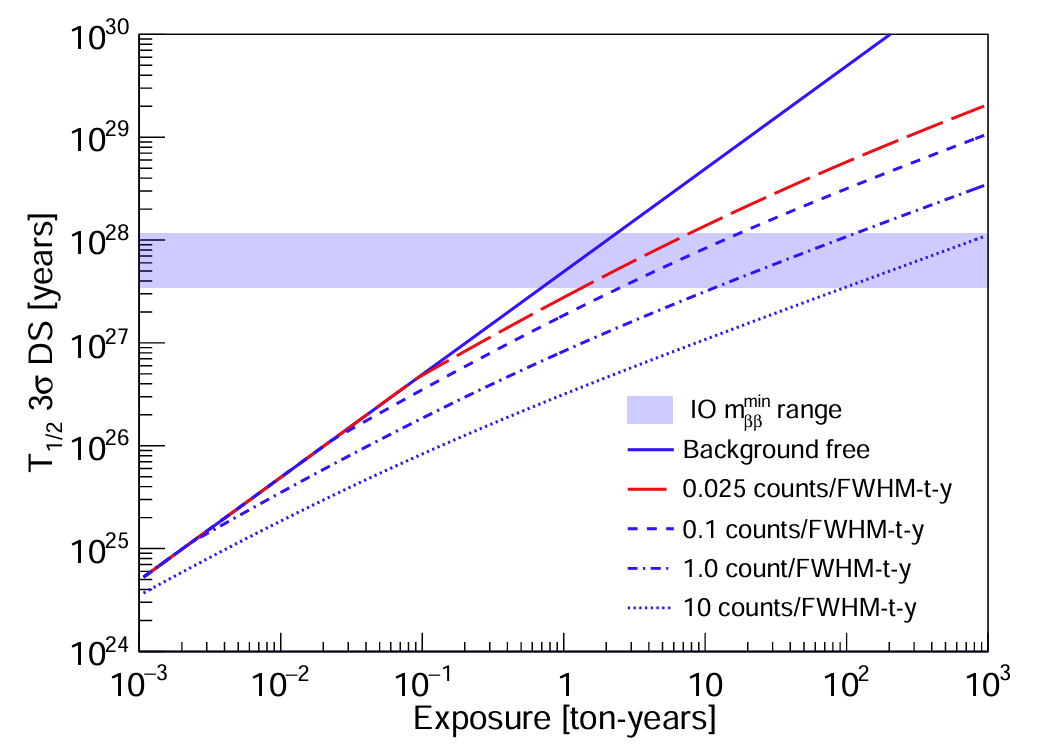}
		}
		\hspace*{-0.7cm}\parbox{\dimexpr\textwidth-0.4cm\relax}{%
			\caption{$3\sigma$ discovery sensitivity as a function of the background level. The red dashed line corresponds to the LEGEND-1000 goal of 10$^{-5}$ cts/(keV kg yr) = 0.025 cts/(FWHM t yr) with FWHM(\qbb) $\sim$ 2.5 keV.}
			\label{discovery_sensitivity}
		}
	\end{figure}
\end{minipage}
\begin{minipage}{0.5\textwidth}
	\begin{table}[H]
		\hspace*{-0.9cm}
		\footnotesize
		\begin{tabular}{lcc}
			\toprule
			& \textbf{LEGEND-200} & \textbf{LEGEND-1000} \\
			\midrule
			Mass [kg] & 200 & 1\,000 \\
			\midrule
			Exposure [kg yr] & 1\,000 & 10\,000 \\
			\midrule
			BI goal [cts/(keV kg yr)] & $2\cdot10^{-4}$ & $10^{-5}$ \\
			\midrule
			Half-life DS (3$\sigma$) [yr] & $\sim 10^{27}$ & >$10^{28}$ \\
			\midrule
			$m_{\beta\beta}$ sensitivity [meV] & 33 -- 89 & 9 -- 24 \\
			\bottomrule
		\end{tabular}
		\hspace*{-0.7cm}\parbox{\dimexpr\textwidth+0.7cm\relax}{%
			\caption{Main specifications and sensitivity goals for LEGEND-200 and LEGEND-1000.}
			\label{tab:legend_project}
		}
	\end{table}
\end{minipage}

\section{LEGEND-200 setup}

The first stage of the experiment, LEGEND-200, started taking data in 2023.
It is operated at Laboratori Nazionali del Gran Sasso (LNGS), located beneath a $\sim$1400 m rock overburden, corresponding to $\sim$3500 m water equivalent, which effectively shields the experiment from most of the cosmic rays showers' content and reduces the cosmic muon flux by six orders of magnitude to a rate of about 1.2 muons/(m$^2$ h).   

The core of the experiment consists of 142.5 kg of high-purity germanium  detectors arranged in 10 vertical strings (101 detectors). Four detector types are deployed: 86.7 kg of Inverted-Coaxial Point-Contact (ICPC) detectors newly designed and produced for LEGEND (41 detectors), 22.1 kg of p-type Point-Contact (PPC) detectors from MAJORANA Demonstrator (26 detectors), 19.0 kg of Broad Energy Germanium (BEGe) detectors from GERDA (28 detectors), and 14.7 kg of semicoaxial (Coax) detectors also from GERDA (6 detectors).
The HPGe detectors employed in the experiment have masses of $\sim$0.5-4 kg and are enriched in $^{76}$Ge up to the 86-92\%, both depending on the detector type; 
they are mounted on scintillating polyethylene naphthalate (PEN) plates using copper rods made from underground-electroformed Cu.

The germanium detectors are operated bare in a 64 m$^3$ liquid argon (LAr) cryostat serving both as a refrigerant ($\sim$88 K) and as an active veto for background events. In fact the scintillation light produced by the interactions taking place in the argon volume is detected by the silicon photomultipliers (SiPM) coupled to two cylindrical curtains of wavelength-shifting polystyrene fibers coated with tetraphenyl butadiene (TPB). The inner barrel, positioned inside the detector string circle, counts 18 readout channels, while the outer barrel counts 40 channels.

A 590 m$^3$ ultra-pure water tank instrumented with 63 photomultiplier tubes (PMTs) surrounds the argon cryostat, serving both as passive shield against muon-induced neutrons and external $\gamma$ radiation and as an active Cherenkov detector for throughgoing muons. The inner surfaces of the water tank are covered with a wavelength-shifting reflector (WLSR) that enhances the light yield and converts scintillation light to 400 nm, matching the PMT sensitivity range.\\

\begin{figure}[H]
	\centering
	\includegraphics[width=0.85\textwidth]{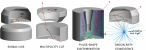} 
	\caption{The leftmost plot shows the expected topology of a signal-like event, with all the energy deposited in a single location (SSE). The following three plots show different topologies -and the corresponding suppression strategies- for background events: the \textit{multiplicity cut} exploits Ge-Ge anticoincidences; the \textit{PSD cut} exploits the correlation between the topology of the energy deposition inside a single Ge detector and the shape of the produced signal; the \textit{LAr veto cut} exploits Ge-LAr anticoincidences.}
	\label{fig:evts_discrimination}
\end{figure}

The electrons emitted in $\beta\beta$ events interact in the germanium within $\sim$1 mm$^3$, so their energy deposition is highly localized: signal events are therefore expected to be Single Site Events (SSE) in the Ge detectors' bulk, which is not the case for many background components. This is the key of most of the background rejection strategies adopted in LEGEND design and analysis (Fig. \ref{fig:evts_discrimination}), with anti-coincidence being a powerful tool to discriminate between signal-like and background-like events.
Additionally, thanks to the geometry of the employed Ge detectors and consequently of the electric field inside them, it is possible to discriminate between SSE and events in which energy is deposited in multiple locations inside a single detector's volume (Multi Site Events, MSE) based on the shape of the resulting signal: this analysis technique is referred to as Pulse Shape Discrimination (PSD).

\section{LEGEND-200 performance}

\begin{figure}[H]
	\centering
	\includegraphics[width=0.49\textwidth]{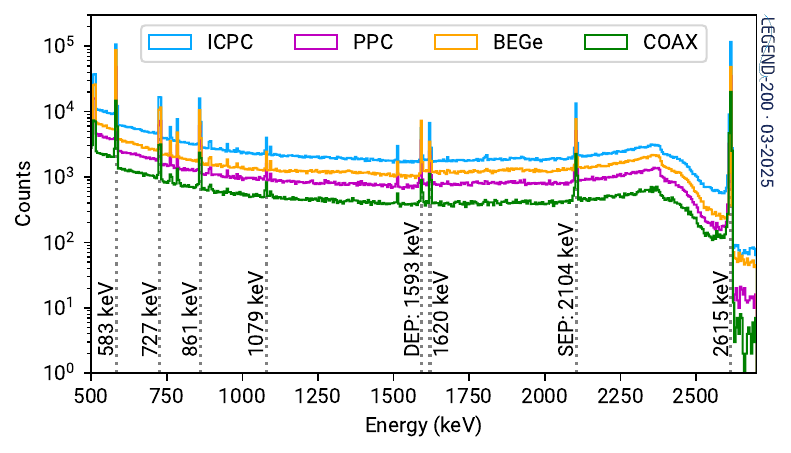}
	\includegraphics[width=0.5\textwidth]{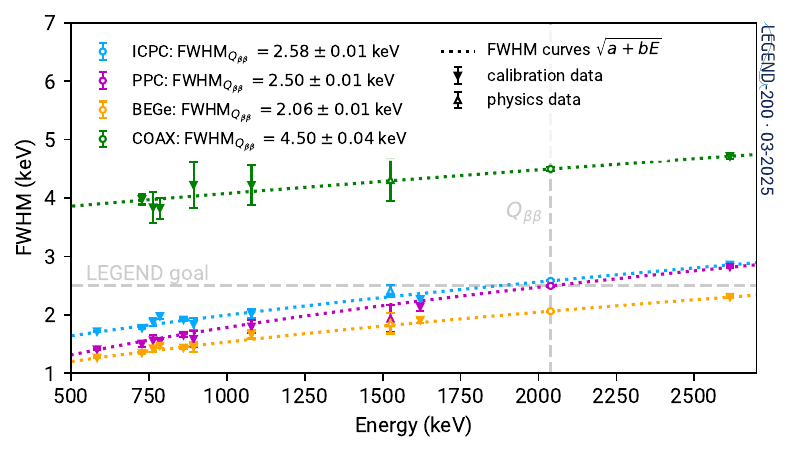}
	\caption{Left: $^{228}$Th calibration spectrum; Right: energy resolution curves.}
	\label{fig:cal_performance}
\end{figure}

From March 2023 to February 2024 LEGEND-200 ran in a stable physics data taking regime, in which the physics data runs are alternated to weekly $\sim$4 hour calibration runs performed with 12 $^{228}$Th sources of $\mathcal{O}$(1 kBq) activity each to evaluate the energy scale, the energy resolution and the PSD performance of the detectors. 
An example of the acquired calibration spectrum is shown in the left panel of Fig.\tild\ref{fig:cal_performance}. The FWHM of the major peaks is calculated to get an estimate of the detectors' energy resolution at different energies; the resulting distribution is fitted with the resolution curve $FWHM(E) = \sqrt{a + bE}$, from which the FWHM(\qbb) is determined, as shown in the right panel of Fig.\tild\ref{fig:cal_performance}. ICPC, BEGe and PPC detectors meet the LEGEND resolution goal of FWHM(\qbb) $\leq$ 2.5 keV. The resolution of Coax detectors, instead, is not compatible with that goal: this was expected from previous GERDA results and will not be an issue in the future as Coax detectors will not be included in the future LEGEND-200 deployments, nor in LEGEND-1000.

An additional tool used to monitor the performance and the stability of the experiment throughout the data taking is the acquisition of pulser signals and forced trigger events. A pulser signal mimicking a $\sim$1 MeV event is injected every 20 s by a pulser module in the front-end electronics of each germanium detector, to monitor the stability of the detectors' and the electronics' response.
Forced triggers are instead signals acquired by each detector every 20 s without any physical event to trigger the acquisition: the content of these events will therefore be either a flat trace (no signal), useful to study the electronics noise, or a random coincidence, the rate of which must also be determined.

For each HPGe detector, the whole collected dataset is split into \textit{partitions}, defined as collections of consecutive runs in which the hardware configuration, the bias voltage, the energy scale and the PSD parameters are stable. 

In the first year of data taking, a total exposure of 85.5 kg yr was collected; from this, 61 kg yr were selected for the first \novbb{} analysis\tild\cite{prl25}.

\section{LEGEND-200 first results}

\begin{figure}[H]
	\centering
	\includegraphics[width=1\textwidth]{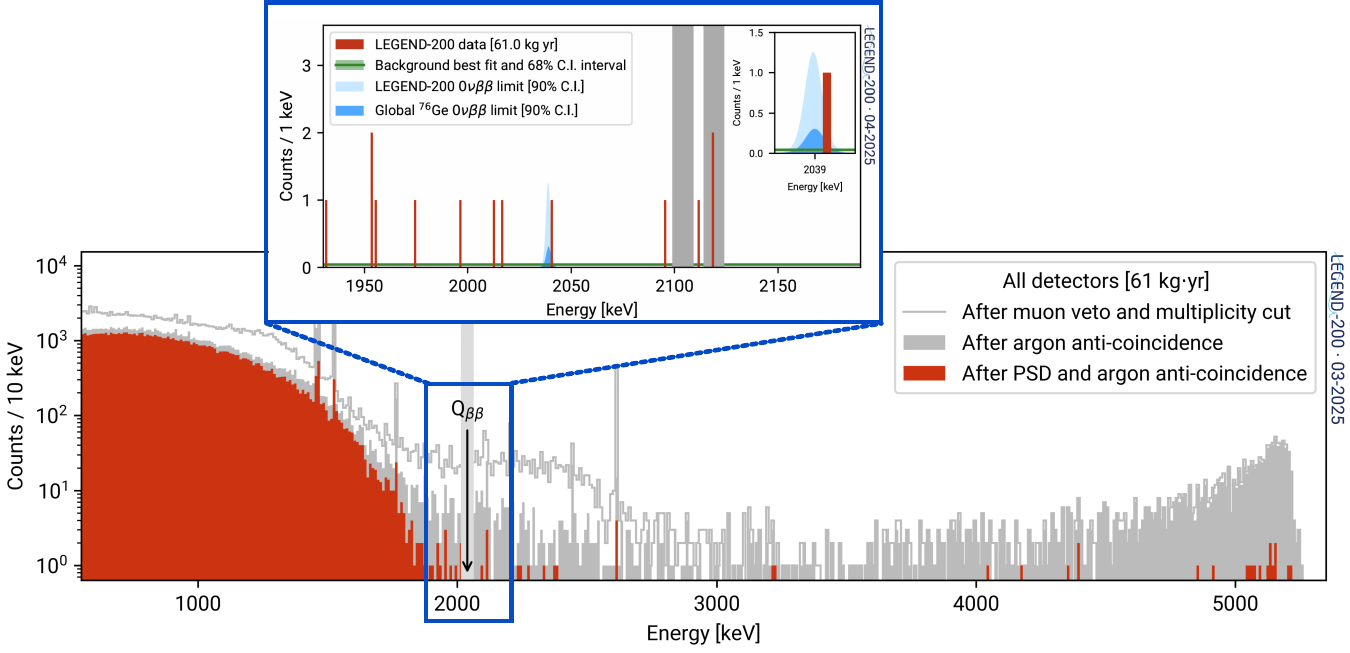} 
	\caption{Energy spectrum acquired by LEGEND-200 (61 kg yr exposure). The efficiencies of the applied cuts are:
	$\epsilon_{muon}(Q_{\beta\beta}),\epsilon_{mult}(Q_{\beta\beta}) > 99.9$\%,
	$\epsilon_{LAr}(Q_{\beta\beta}) \sim 93$\%,
	$\epsilon_{PSD}(Q_{\beta\beta}) \sim 76-85$\%.
	Inset: unblinded \novbb{} analysis window.}
	\label{fig:final_spectrum}
\end{figure}

The energy spectrum acquired by LEGEND-200 during its physics data taking is shown in Fig.\tild\ref{fig:final_spectrum}. Pulser signals, forced triggers, and all the events that do not pass the quality cuts are not included in the plotted spectrum.
All the data selection routines are optimized and applied in a strictly blind configuration, with the blinded window being $Q_{\beta\beta} \pm 25$ keV.

The white histogram contains all the events that pass the \textit{HPGe multiplicity cut}, removing all the events in which multiple signals are acquired in coincidence in different germanium detectors, and the \textit{muon veto cut}, removing all the events for which a muon is detected by the water tank's PMTs within $\pm1.75 \  \upmu $s from a HPGe trigger. This spectrum includes all the expected features: the \twovbb{} distribution dominating below \qbb{}, $\gamma$ lines and continuum from $^{40}$K, $^{238}$U, $^{232}$Th and $^{42}$K, the last being progeny of the $^{42}$Ar present in the liquid argon (below $\sim$3 MeV), some $\beta$ contribution from $^{42}$K (Q-value = 3.53 MeV) and the $\alpha$ population from the $^{238}$U chain at higher energy (4-5 MeV). 
When compared to radioassay-based background predictions, though, the acquired spectrum appears to have a higher background level, also in the region around \qbb{}; this background excess appears to be successfully mitigated by the LAr veto and PSD cuts, but it was nevertheless the main subject of investigation in the past year, after the stop of the physics run employed in this analysis.

The gray histogram contains only the events that, in addition, pass also the \textit{LAr veto cut}: it removes all the events for which a sizable amount of energy is released in the liquid argon in a [-1,5] $\upmu$s coincidence window from a HPGe trigger. For this analysis the veto condition is either that $\geq$4 photoelectrons in total are collected by the SiPMs or that $\geq$4 SiPMs acquire a signal (above threshold).

Finally, the red histogram contains only the events passing also the \textit{PSD cut}. For this analysis the PSD cut is based on two parameters: A/E, which is defined as the ratio between the maximum amplitude of the current pulse (A) and the energy of the event (E), and LQ (late charge), defined as the area above the charge signal after the time at which it reaches the 80\% of its total amplitude.
For ICPC detectors produced by Mirion and BEGe detectors, a double sided cut is applied to the A/E parameter: the low side cut, optimized to achieve a 90\% survival fraction of the SSE-dominated $^{208}$Tl double escape peak (DEP) in calibration data, effectively suppresses MSE and events happening close to the n+ surface of the detector, both characterized by a lower amplitude of their current pulse; the high side cut, fixed at a value of 3, effectively suppresses events happening close to the p+ contact ($\alpha$ particles), which are typically fast signals (higher amplitude of the current pulse).
For PPC detectors and ICPC detectors produced by Ortec, due to their different geometry translating to different properties of their electric fields and consequently different pulse shapes, the applied PSD cut combines instead the low side A/E cut for MSE to an LQ $<$ 3 cut to suppress surface events.
Finally for Coax detectors the standard PSD parameters are not effective, due again to the impact of the geometry on the pulse shape features: an Artificial Neural Network (ANN)-based PSD cut is used instead. 
The energy spectrum after all cuts is dominated by the \twovbb{} distribution below \qbb{}, while only few residual counts survive near and above it.

\vspace*{-0.2cm}

Two separate unblindings were performed for two subsets of the acquired data, which were expected to show different background levels.
The \texttt{golden dataset} includes 48.3 kg yr of exposure from BEGe, PPC and ICPC produced by Mirion, while the \texttt{silver dataset} includes 12.7 kg yr of exposure from COAX and ICPC produced by ORTEC.

The inset of Fig\tild\ref{fig:final_spectrum} shows the unblinded spectrum after all cuts in the energy range [1930-2190] keV considered for the statistical analysis aimed at the calculation of the background index (BI) and of the half-life limit. Two known gamma lines, $[2104\pm5]$ keV from $^{208}$Tl and $[2119\pm5]$ keV from $^{214}$Bi, are excluded from the analysis window.
A total of 11 events survive all the cuts in the analysis window, 7 coming from the \texttt{golden dataset} and 4 from the \texttt{silver dataset}.

The frequentist analysis finds no evidence of a \novbb{} signal and sets a lower limit on its half-life to $T_{1/2}^{0\nu} > 0.5 \cdot 10^{26}$ yr at 90\% confidence level (CL), with a median exclusion sensitivity of $1.0\cdot10^{26}$ yr.
The background index is $0.5_{-0.2}^{+0.3} \cdot 10^{-3}$ counts/(keV kg yr) in the \texttt{golden dataset} and $1.3_{-0.5}^{+0.8} \cdot 10^{-3}$ counts/(keV kg yr) in the \texttt{silver dataset}.  
A combined GERDA (127.2 kg yr) + MAJORANA Demonstrator (64.5 kg yr) + LEGEND-200 (61 kg yr) fit is ultimately performed: again no evidence of a \novbb{} signal is found and a lower limit to its half-life is set to $T_{1/2}^{0\nu} > 1.9 \cdot 10^{26}$ yr at 90\% CL, with a median exclusion sensitivity of $2.8\cdot 10^{26}$ yr, the best achieved among \novbb{} decay experiments to date. The corresponding Bayesian analysis provides compatible results, both for the LEGEND-only and for the combined fit. The resulting half-life limit can be used to place a limit on the neutrinos effective Majorana mass: using a range of nuclear matrix elements from theoretical calculations, $m_{\beta\beta} <$  75–200 meV.

\section{Conclusion}

From March 2023 to February 2024 LEGEND-200 completed its first year of physics data taking, collecting a total exposure usable for the \novbb{} analysis of 61 kg yr. The performed frequentist analysis finds no evidence of a \novbb{} signal and sets a lower limit on its half-life to $T_{1/2}^{0\nu} > 0.5 \cdot 10^{26}$ yr at 90\% CL, with a median exclusion sensitivity of $1.0\cdot10^{26}$ yr; the joint GERDA + MAJORANA Demonstrator + LEGEND fit sets the lower limit to $T_{1/2}^{0\nu} > 1.9 \cdot 10^{26}$ yr at 90\% CL, with a median exclusion sensitivity of $2.8\cdot 10^{26}$ yr, the best achieved to date.\\

The quality of the data acquired in this first year is high and mostly compliant to the expectations and goals; yet a background excess with respect to radioassay-based predictions was observed in the energy spectrum before analysis cuts, impacting also the region of interest around \qbb{}. 
For this reason an extensive screening and cleaning campaign was undertaken, with the goal of identifying and mitigating the background sources responsible for this excess.
Starting from April 2025, the array was redeployed, with some major upgrades with respect to the original configuration, besides the cleaning of the components. A total of 137.5 kg of germanium detector is currently operated, including only the best detectors from the first deployment: Coax and PPC detectors are no more deployed, while only ICPC and BEGe detectors are operated, including numerous newly produced, high mass, ICPC detectors. 
A first analysis of the data acquired in the new configuration suggests that a sizable background reduction was achieved; further details and more quantitative results will be provided once the acquired statistics increases and more complete analysis are performed.

\end{document}